\documentclass[dvips]{elsarticle}
\usepackage{graphicx}

\newcommand{\dt }   {\partial_t}
\newcommand{\dth }  {\partial_\theta}
\newcommand{\dr }   {\partial_r}
\newcommand{\gradpar }{\nabla_\parallel}

\newcommand{\rhost} {\rho_{\displaystyle *}}

\def\BE{\begin{equation}}
\def\EE{\end{equation}}
\def\BEA{\begin{eqnarray}}
\def\EEA{\end{eqnarray}}

\def\vbar{|}

\def\vB{{\bf B}}
\def\vb{{\bf b}}
\def\vx{{\bf x}}

\def\vuz{{\hat{\bf z}}} 

\def\vnabla{{\bf \nabla}}

\begin{document}


\title{An alternative approach to field-aligned coordinates for plasma turbulence simulations}

\author[CAD]{M. Ottaviani}
\ead {Maurizio.Ottaviani@cea.fr}
\address[CAD]{CEA, IRFM, F-13108 Saint-Paul-lez-Durance, France.}
\date{\today}

\begin{abstract}
Turbulence simulation codes can exploit the flute-like nature of plasma turbulence to reduce the effective number of degrees of freedom necessary to represent fluctuations. This is achieved by employing magnetic coordinates of which one is aligned along the magnetic field. In the most common implementation, positions along the field lines are identified by the poloidal angle. 

In this work, an alternative approach is presented, in which the position along the field lines is identified by the toroidal angle. It will be shown that this approach has several advantages. Among these, periodicity in both angles is retained. 

This property allows moving to an equivalent representation in Fourier space with a reduced number of toroidal components. It will be shown how this duality can be exploited to transform conventional codes that use a spectral representation on the magnetic surface into codes with a field-aligned coordinate.

It is also shown that the new approach can be generalised to get rid of magnetic coordinates in the poloidal plane altogether, for a large class of models. 

Tests are carried out by comparing the new approach with the conventional approach employing a uniform grid, for a basic ion temperature gradient (ITG) turbulence model implemented by the two corresponding versions of the ETAI3D code. 

These tests uncover an unexpected property of the model, that localized large parallel gradients can intermittently appear in the turbulent regime. This leaves open the question whether this is a general property of plasma turbulence, which may lead one to reconsider some of the usual assumptions on micro-turbulence dynamics.
\end{abstract}


\maketitle
 
\section{Introduction}

Microturbulence in tokamaks has the property that gradients parallel to the magnetic field are substantially smaller than gradients in the perpendicular direction. This is the consequence of low-frequency dynamics combined with a very efficient transport along field lines. On can assume for example that the typical evolution rate is of the order of the diamagnetic frequency 
$\omega_{\displaystyle{*}} = (c T_e / e B) k_\theta / L$, 
where $L$ is a macroscopic (temperature or density) gradient scale length, and that parallel transport occurs at the rate given by free ion streaming (which can be considered typical of weakly collisional plasmas) $\gamma_\parallel \approx v_{thi} k_\parallel$. Balancing these two rates one obtains

\begin{equation}
\label{ratio}
k_\parallel / k_\theta \sim \rho_s/L \sim \rho_{\displaystyle{*}}
\end{equation}
Codes have been developed that take advantage of this property to reduce the number of grid points to maximize code efficiency~\cite{Ham93,Dim93,Scott98,Scott01}. Moreover, using a coarse grid in the parallel direction reduces the time scale constraint given by the numerical stability condition, making the use of explicit schemes less penalising. The combination of these two factors gives to codes that are able to exploit this property an overall efficiency much superior than the one of conventional codes that employ a uniform grid and implicit schemes to deal with the parallel dynamics.

This work explores ways to implement coarse gridding in the parallel direction by making use of field-alignment. The initial motivation was the conversion of a certain class of conventional codes (spectral in the angles) to a more efficient representation.

It turns out that there are just two practical ways to achieve this goal, which differ in the way one labels points along the field line. In the quasi-totality of codes that make use of field alignment, one employs a coordinate transformation such that the parallel direction turns out to be effectively labelled by the {\sl poloidal angle}. A second coordinate is needed to represent the fine structure of turbulent fields on a magnetic surface; this is effectively the toroidal angle. The alternative approach proposed in this work reverses the roles, by effectively using the toroidal coordinate to label the position of a point along the field lines, in a way that will be explained in detail in the following sections. The toroidal angle was previously employed as ``parallel'' coordinate in the ``quasiballooning coordinates'' construction of Dimits~\cite{Dim93}, which differs in important ways from the present work, as discussed later.

It turns out that the approach pursued in this work has a number of advantages.
a) It allows a simpler implementation of nonlinearities for a large class of models.
b) It retains the original representation of functions in the poloidal plane.
c) It can cope with X-point configurations including the open field line region.
d) Functions are represented on a regular grid in real space. This allows moving to an equivalent representation in Fourier space with a reduced number of components. This duality can be exploited to transform conventional codes that use a spectral representation on the magnetic surface into codes with a field-aligned coordinate.
e) Finally, as a further development, the new approach can be generalised, for a large class of models, to get rid of magnetic coordinates in the poloidal plane altogether. Only the toroidal angle is needed to identify positions along the field lines to compute parallel gradients.

This work is organised as follows. Section~\ref{review} reviews the main aspects of field alignment techniques. The new approach will then be derived in Sec.~\ref{new-approach}, giving first an intuitive description and then a mathematical justification.
Section~\ref{code} will show how the spectral ITG code ETAI3D~\cite{MO,FO} can be converted to the new system with minimal programming. 
Section~\ref{tests} is devoted to the comparison between the old and the new version of the code. These tests uncover an unexpected property of the model. While gradients parallel to the magnetic fields are typically small, as expected as a consequence of the flute property (\ref{ratio}), localized sharp parallel gradients can appear in the turbulent regime. This is a property of the model (not of the method) which can have important consequences in terms of convergence.
Generalisation to a system not using magnetic surfaces is described in the Appendix.

\section{Review\label{review}}

\begin{figure}
\centering
	\includegraphics[width=8cm]{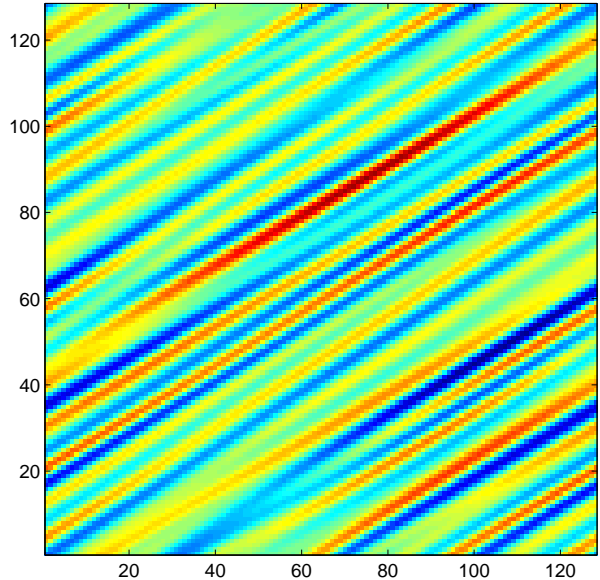}
	\caption{Chart of electric potential fluctuations on a given magnetic surface}
\label{fig-isolines}
\end{figure}

The problem posed by the two-scale nature of plasma micro-turbulence is illustrated in Fig.~\ref{fig-isolines}. 

\begin{figure}
\centering
	\includegraphics[width=8cm]{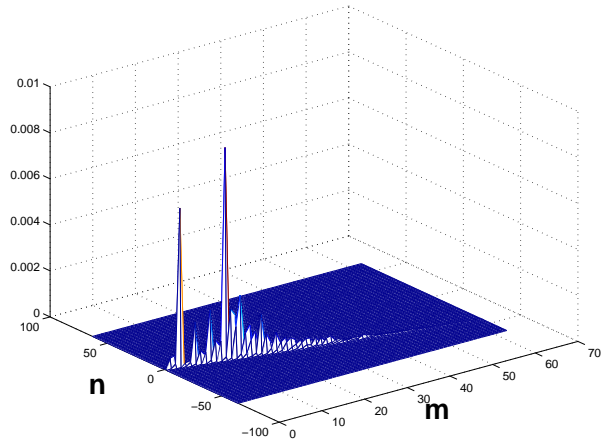}
	\caption{Temperature fluctuations spectrum at a given radial position}
	\label{fig-spectrum}
\end{figure}

This figure shows a chart of the turbulent electrostatic potential $\Phi$ at a fixed radial position, that is, on a given magnetic surface, as a function of the toroidal angle $\varphi$ (abscissa) and poloidal angle $\theta$ (ordinate). Details of how this figure was obtained will be given later. One can clearly observe a direction along which structures tend to align along the lines $\theta=\varphi /q(r) + {\rm const.}$ , which correspond to the local magnetic field direction, $q(r)$ being the safety factor at the given radial location $r$. It is clear that turbulent structures have generically small parallel gradients.
Carrying out a Fourier transform to mode number space $(n,m)$ one obtains the spectrum $\vbar\Phi_{m,n}\vbar^{2}$ given in Fig.~\ref{fig-spectrum}. 

As expected, most of the turbulent energy is concentrated along the lines $m=nq$. It is apparent how much computer resources are wasted by a representation of the turbulent fields that uses a grid with uniformly small spacing in the angles, as well as by its corresponding representation in mode number space. Clearly, a coordinate system that would exploit the property of small parallel gradients would constitute a substantial advantage.

For the sake of simplicity, only magnetic configurations with circular concentric magnetic surfaces will be considered in this work. Usual radial  and angular coordinates ($r,\theta,\varphi$) are employed. Information on the local magnetic field direction is contained in the safety factor $q(r)$. The derivative along the magnetic field is embodied in the parallel derivative operator:
\begin{equation}
\label{par-der-defin}
\nabla_{\parallel} = \frac{\partial}{\partial \varphi}  + \frac{1}{q(r)} \frac{\partial}{\partial \theta}.
\end{equation}
This operator is proportional to the actual parallel derivative operator by an $r$-dependent factor.
The first implementation of field aligned coordinates in a plasma turbulence code can be traced back to thinking in the early Nineties~\cite{CKS} that led to Ref. \cite{Ham93}.
Consider the transformation:
\begin{eqnarray}
&& \xi=\varphi - q(r) \theta \nonumber
\\
&& s=\theta \label{coord-tr-old}
\\
&& \rho=r \nonumber
\end{eqnarray}
Derivatives with respect to the original variables are given by
\begin{eqnarray}
&& \frac{\partial}{\partial r} = \frac{\partial}{\partial \rho} - q \prime (r) \theta \frac{\partial}{\partial \xi} \nonumber
\\
&& \frac{\partial}{\partial \varphi} = \frac{\partial}{\partial \xi} \label{coord-der-old}
\\
&& \frac{\partial}{\partial \theta} = \frac{\partial}{\partial s} - q(r) \frac{\partial}{\partial \xi}
\nonumber
\end{eqnarray}
While the parallel derivative is given by
\begin{equation}
\label{par-der-tr-old}
\nabla_\parallel = \frac{1}{q(r)} \frac{\partial}{\partial s}
\end{equation}
Since parallel derivatives are expected to be small, a small number of points along $s$ is needed to represent a function adequately. In the light also of Eq.~\ref{coord-tr-old}, one can equivalently say that a small number of poloidal points is needed to represent a given function. One also observes that the poloidal derivative is given by a linear combination of a (slow) derivative along $s$ and of a (fast) derivative along $\xi$. Since the latter is a derivative taken along $\varphi$, one can say that the transformation (\ref{coord-tr-old}) decomposes the poloidal derivative into a parallel component and a toroidal component.
Since the parallel variable is coarse, in this representation all the information on the fine structure of turbulence is necessarily carried by the toroidal angle. Also, in actual code implementation, the derivative along $s$ is usually dropped, approximating the poloidal derivative by
$\partial / \partial \theta \approx q(r) \partial / \partial \xi$. In linear theory, for a given toroidal mode $n$, one can Fourier-transform along $\varphi$ ($\xi$), which gives $\partial / \partial \theta \approx i n q(r) $. One can see that transformation (\ref{coord-tr-old}) is equivalent to the ballooning transformation\cite{CHT}, as already remarked in \cite{KW}. 

This coordinate transformation, as it is, has some drawbacks. First, one notices that the new coordinate $s$ is not periodic. Care must be taken, when setting boundary conditions at the end points of an $s$ line, that the original double-periodicity of the given magnetic surface is enforced~\cite{Scott98}. Non-compliance with this constraint would lead to spurious solutions and dubious results. 
The second problem is the consequence of the term, proportional to $\theta$, appearing in the expression of the radial derivative Eq.~\ref{coord-der-old}. This term, familiar from the ballooning transformation, leads to mixed-derivatives of increasing weight as one moves away from $\theta=0$, when computing certain operators as the Laplacian. These terms are the consequence of $\theta$-dependent non-diagonal metric coefficients in the new coordinate system. Although mathematically correct, they pose a numerical challenge since their numerical treatment can introduce artificial inhomogeneities in the poloidal direction even for system possessing poloidal symmetry. This problem is cured by the shifted-metric technique\cite{Scott01} as described below.
A third problem with the old coordinate transformation is that it cannot deal with the separatrix, since, there, the safety factor becomes infinite. This problem is automatically solved with the new coordinate transformation proposed in this work. 

The shifted-metric approach introduced in \cite{Scott01} consists in sectioning the toroidal manifold given by the magnetic surface into a number of parts, $N_\theta$, each having its own coordinate system differing one from the other by a shift in the origin. Here a slightly different version is given, which makes use of overlapping poloidal sectors. Consider the set of transformations 
\BEA
&& \xi=\varphi - q(r)( \theta - \theta_k) \nonumber
\\
&& s_k=\theta-\theta_k \label{coord-tr-old-shift}
\\
&& \rho=r \nonumber
\EEA
where $\theta_k=\Delta \theta k$ with $\Delta \theta =2 \pi / N_\theta$ and $k=0,N_\theta-1$.
A given sector is defined by the rectangle such that $0 \le \varphi \le 2 \pi$ and $\theta_k - \Delta \theta \le \theta \le \theta_k + \Delta \theta$.
Derivatives with respect to the original variables are now given by
\BEA
&& \frac{\partial}{\partial r} = \frac{\partial}{\partial \rho} - q \prime (r) (\theta-\theta_k) \frac{\partial}{\partial \xi} \nonumber
\\
&& \frac{\partial}{\partial \varphi} = \frac{\partial}{\partial \xi} \label{coord-der-old-shift}
\\
&& \frac{\partial}{\partial \theta} = \frac{\partial}{\partial s} - q(r) \frac{\partial}{\partial \xi} \nonumber
\EEA
One observes that, for each poloidal sector $k$, the additional term that leads to metric distortion vanishes at $\theta=\theta_k$. If now one uses exactly the same number of sectors as the number of points along $\theta$ needed to achieve the required resolution, one realises that it is necessary to compute perpendicular operations only at points where no metric distortion occurs. The computation of parallel derivatives requires the knowledge of the functions to be derived at the end points $s_k=\pm \Delta \theta$  of an $s_k$ line. Whereas one could use collocation points at these ends, this would still leave the question of relating values of the function at these points with the values at the $s_{k \pm 1}=0$ lines of the neighbouring $k \pm 1$ sectors. In practice, the solution adopted in \cite{Scott01}, which will be maintained here, is to assign values of a function only on the $s_k = 0$ lines, and, for each $k$, to compute the values at the end points $s_k = \pm \Delta \theta$ by interpolation of the values at the $s_{k \pm 1} = 0$ lines of the neighbouring $k \pm 1$ sectors. Notice that the accuracy needed to compute the values of a function at the interpolation points is automatically assured by the high resolution needed to describe a function on a given $s_k=0$ line.

\section{New approach to field-aligned coordinates \label{new-approach}}

\begin{figure}
\centering
	\includegraphics[width=8cm,height=8cm]{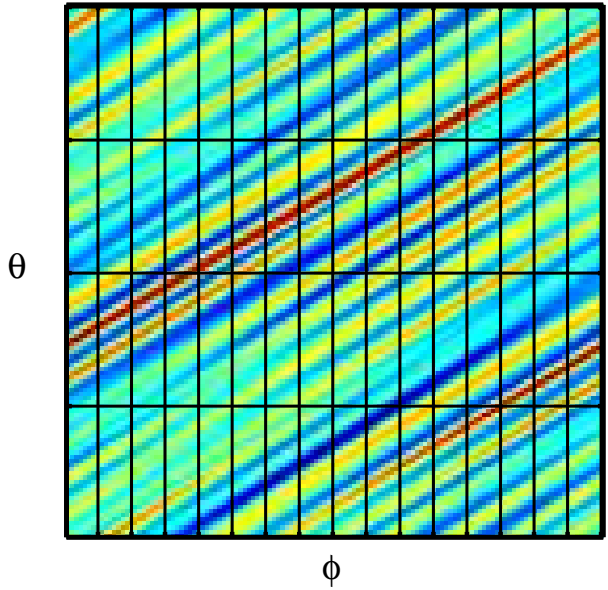}
	\caption{Mesh on a magnetic surface with reduced resolution in $\theta$ (old approach)}
	\label{fig-mesh-a}
\end{figure}

One starts with the observation that the net result of the shifted metric procedure is that functions are now defined on a regular grid in the former $(\varphi,\theta)$ space, with the advantage that the discretisation is now coarser in $\theta$ than it would be in a conventional meshing with uniformly small spacing in both directions. The coordinate $\theta$ labels the position along the field lines. Also, this mesh still has the good property of being periodic in both directions, which are effectively the original angles.
The situation is described in Fig.~\ref{fig-mesh-a}. 

\begin{figure}
\centering
	\includegraphics[width=8cm,height=8cm]{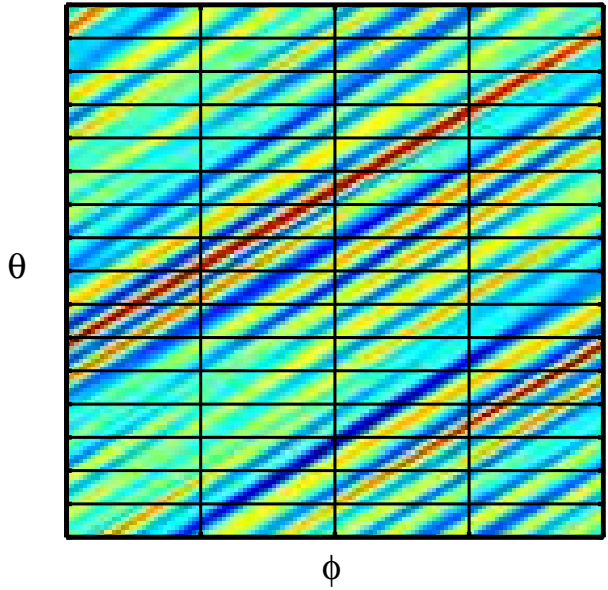}
		\caption{Mesh on a magnetic surface with reduced resolution in $\varphi$ (new approach)}
\label{fig-mesh-b}
\end{figure}

One also remarks that, as a consequence of the flute property, one can reduce the resolution in any chosen direction, provided that the
information on the fine structure of the turbulent fields can be carried by the variation in any other direction.
One then realises that there is just another alternative to this meshing, which preserves the good property of double periodicity. It is given by switching the roles of coarse/fine mesh between the poloidal/toroidal angles, as shown in Fig.~\ref{fig-mesh-b}. 

This latter choice has a coarse toroidal grid. This implies that it is now the toroidal angle that must be used to describe variations along the field lines.
This can be justified mathematically by sectioning the toroidal manifold in $N_\varphi$ overlapping toroidal sectors, each with its own set of field-aligned coordinates given by the family of transformations 
\BEA
&& \xi=\theta - \frac{1}{q(r)} (\varphi - \varphi_k) \nonumber
\\
&& s_k=\varphi-\varphi_k \label{coord-tr-new}
\\
&& \rho=r \nonumber
\EEA
where $\varphi_k=\Delta \varphi k$ with $\Delta \varphi =2 \pi / N_\varphi$ and $k=0,N_\varphi-1$.
Derivatives with respect to the original variables are now given by
\BEA
&& \frac{\partial}{\partial r} = \frac{\partial}{\partial \rho} - \frac{q \prime (r)}{q^2(r)} (\varphi - \varphi_k) \frac{\partial}{\partial \xi} \, , \nonumber
\\
&& \frac{\partial}{\partial \varphi} = \frac{\partial}{\partial s} - \frac{1}{q(r)} \frac{\partial}{\partial \xi} \, , \label{coord-der-new}
\\
&& \frac{\partial}{\partial \theta} = \frac{\partial}{\partial \xi} \, , \nonumber
\EEA
whereas the parallel derivative is now simply given by
\begin{equation}
\label{par-der-tr-new}
\nabla_\parallel =  \frac{\partial}{\partial s} \, .
\end{equation}
On any given $\varphi_k=0$ line, the derivatives are given by the same expressions as the original derivatives. 
Parallel derivatives are computed as before by interpolation at the end points of the $\varphi_k$ lines.

This new system has additional advantages. The fact that the transformation involves $1/q(r)$ rather than $q(r)$ allows one to take the limit of $q \rightarrow \infty$ without problems; this permits to treat a system whose magnetic equilibrium possesses a separatrix. 
Also, the poloidal derivative is simply given by a derivative along $\xi$, while the radial derivative is a simple derivative along $\rho$ (at $\varphi=\varphi_k$). 
This permits to implement key operators like the Laplacian or perpendicular Poisson brackets in a manner strictly equivalent to the one of the initial coordinate system.
This fact will be exploited in Sec.~\ref{tests} to carry out a comparison between mathematically equivalent codes that employ the conventional and the field-aligned coordinate systems.

As mentioned in the Introduction, the approach presented here has in common with Ref.~\cite{Dim93} the use of the toroidal angle to keep track of the position along the field lines. However, there are several important differences. Ref.~\cite{Dim93} uses, for each radial point, an (optimal) rational approximation to $1/q(r)$ in what would be Eq.~\ref{coord-tr-new}. As a result, parallel derivatives expressed in terms of the new variables still contain a contribution from the derivative along the fast variable, albeit with a small coefficient. Moreover, points are collocated along the new almost-parallel direction, without use of the shifted-metric idea, and as a consequence one looses periodicity in one direction. Perpendicular operators like the Laplacian need to be computed with interpolation, unlike in the present work, where interpolation is necessary only to compute parallel derivatives.

Finally, the approach presented here can be extended to avoid using magnetic coordinates in the poloidal plane, as sketched in the Appendix.

\section{Implementation of field-alignment in the ETAI3D code \label{code}} 

In order to test the new approach, one has first to implement it in some code that solves a three-dimensional micro-turbulence model with a sheared magnetic equilibrium as minimal features. It was found convenient to start from an existing code, ETAI3D, which 
solves a model of ITG turbulence described by the following equations:
\begin{eqnarray}
&& \frac{d}{dt} w  - 2 \;\varepsilon\;\omega_d \;(\Phi + \tau w + T_i) + A \gradpar v_\|
 = 
\nabla (D_w \nabla w)\,, 
\label{w}
\\
&& \frac{d}{dt} v_\|  - 4\; \tau\; \;\varepsilon\; \omega_ d\; v_\| + A \gradpar (\Phi + \tau w + T_i)\,
 = 
\nabla (D_v \nabla v_\|),
\label{v}
\\
&& \frac{d}{dt} T_i  - 2 \;\tau \;\varepsilon
 \;\omega_d  \left((\Gamma -1) (\Phi + \tau w ) + (2\,\Gamma -1) T_i \right) +  \nonumber \\
&& + A \;\tau\; (\Gamma - 1) \gradpar v_\| 
 = 
- A \gamma_L\;{\tau}^{1/2}\; |\gradpar|\, T_i + \nabla (D_T \nabla T_i) \,.
\label{T}
\end{eqnarray}

Here
$w$ is the ion guiding centre density, $v_\|$ the ion parallel velocity and
$T_i$ the ion temperature. $\Phi$ is the electrostatic potential,
which is related to $w$ by
$w = \Phi - <\Phi> - \rhost^2 \nabla_\perp^2 \Phi$, where $<.>$ denotes the flux surface average.
The equations are implemented on a cylinder $(r,\theta,\varphi)$ where $\varphi$ is the direction of the principal (would-be toroidal) magnetic field.

The operators in Eqs.~(\ref{w}-\ref{T}) are: a) the total time derivative in the
ExB velocity field $\vec v_E$, $\frac{d}{dt} f = (\dt  + \vec v_E \cdot \nabla) f \equiv \dt f + [\Phi, f ] \;$, where the Poisson bracket,
in cylindrical coordinates, is $[\Phi,f] = \dr \Phi \frac{1}{r}\dth f -  \dr f \frac{1}{r}\dth \Phi $; b) the curvature operator $\omega_d = \sin \theta \,\dr + \frac{1}{r} \cos \theta\,\dth $; c) the parallel derivative operator
$\gradpar = \partial_\varphi + \frac{1}{q} \dth$, with $q$ the safety factor. 

The parameters of the model are the ion sound Larmor radius normalized to the minor radius
 $\rhost = {\rho_s}/{a}$, the inverse aspect ratio $\varepsilon = a/R$, $A = \varepsilon/\rhost$, a constant $\Gamma=5/3$, the temperature ratio $\tau= T_{i0} / T_{e0}$, the small perpendicular transport coefficients 
$D_w$, $D_v$ and $D_T$, set to damp the small scales, and $\gamma_L=1$ is the coefficient of the parallel heat flux in Eq.~(\ref{T}), modelled by a Landau-fluid closure 
of the Hammett-Perkins type~\cite{HP}, as appropriate in the weakly collisional, long mean-free-path limit.

Lengths are normalised to the minor radius $a$, times to the Bohm unit $t_{Bohm} = a^2/\chi _{Bohm}$, with $\chi _{Bohm} = \rho_s c_s$, the parallel velocity to the ion sound velocity $c_s= \sqrt{{T_e}/{m_i}}$, the ion temperature to $T_e = T_{e0}$ (assumed spatially constant) and the potential to $T_e /e$.

Note that this model, involving only three scalar fields, is substantially simpler than ITG models implemented in more modern codes that aim to reproduce a good agreement with observations. However it is adequate for testing purposes in the context of the present work.

The fundamental variables evolved by the code are the Fourier components of the fields in the angular variables, truncated to given maximum poloidal and toroidal mode numbers, and defined at given points on a discrete radial grid. Derivatives are computed by second order finite differences in the radial direction and by direct multiplication by the corresponding wave numbers in the angular directions. Quadratic nonlinearities are computed with the pseudo-spectral method, by transforming first to real space, computing the products there, and coming back to Fourier space in the end. A de-aliasing technique as given in Ref.~\cite{OP} is employed to ensure that the pseudo-spectral procedure to calculate nonlinearities is strictly equivalent to a direct convolution truncated at the maximum wave numbers. Time advancing makes use of a splitting technique where the reactive part of the model (nonlinearities and curvature) is advanced by a leap-frog scheme, and the diffusive and parallel operations by implicit schemes that keep the algorithm to overall second order accuracy.

In this standard spectral representation, the parallel derivative operator~(\ref{par-der-defin}) is represented by $n+m/q(r)$.  
As already shown in Fig.~\ref{fig-spectrum}, this type of spatial discretisation is rather inefficient. 

Consider now the question of the implementation of the parallel derivative of a given function $F(r,\theta,\varphi)$ as obtained by the field-alignment transformation~(\ref{coord-tr-new}). At any given point in real space, one can use a second order centred finite difference (FD) expression of (\ref{par-der-tr-new}), computed with a suitable spacing $\Delta s$ along the coordinate $s$, while holding $\xi$ and $\rho$ constant. In the original coordinate system, the FD expression of the parallel derivative then reads:
\BEA
&& \nabla_\parallel F = \frac {F(s + \Delta s) - F(s-\Delta s)}{2 \Delta s} = \nonumber 
\\
&& \frac{F(r,\theta + \Delta \varphi /q(r), \varphi + \Delta \varphi) -
F(r,\theta - \Delta \varphi /q(r), \varphi - \Delta \varphi)}{2 \Delta \varphi} \label{par-der-FD}
\EEA
where the fact that $\Delta s = \Delta \varphi$ has been used. We note again that   
values of the function at the end points $(\varphi \pm \Delta \varphi,\theta \pm \Delta \varphi/q)$ along the field lines need to be computed by interpolation. For this, a good knowledge of the function along the poloidal direction (lines of constant $\varphi$) is necessary. That is, one must have in general $\Delta \theta << \Delta \varphi$ to resolve the turbulence scale length in the poloidal direction, whereas $\Delta \varphi$ needs not be very small.

Since one works with a doubly periodic domain in the angles, interpolation is conveniently carried out by transforming to Fourier (mode number) space.
The two representations have the same information content. This implies that if a small number of toroidal points gives an adequate representation of a function, the same (small) number of toroidal modes is needed, provided that 
the resolution in the poloidal direction is adequate.
 
Consider the Fourier representation of a function $F$, written in terms of its components $F_{mn}$. 
\begin{equation}
F (r,\theta,\varphi) = \sum_{m,n} F_{mn}(r) e^{i (m \theta + n \varphi)}
\end{equation}
Once $F$ is known on the grid nodes in real space, this expression can be used to obtain approximate values of the function at all points in space. 
The FD expression of the parallel derivative (\ref{par-der-FD}) becomes
\begin{equation}
\label{par-der-Fourier}
\nabla_\parallel F = \sum_{m,n} \frac{\sin [(n+m/q(r)) \Delta \varphi]} {\Delta \varphi}F_{mn}(r) e^{i (m \theta + n \varphi)}
\end{equation}

One then observes that the expression
\begin{equation}
\label{k-par-sin}
\frac{\sin [(n+m/q(r)) \Delta \varphi]} {\Delta \varphi}
\end{equation}
is the representation of the parallel derivative operator in mode number space, in its approximate FD form.
One notes that for small enough values of $\Delta \varphi$, such that $\max [(n+m/q(r)) \Delta \varphi] <<1 $ this expression reduces to the usual one $n+m/q(r)$. However, from the previous discussion, one expects that expression (\ref{k-par-sin}) can be used also with substantially larger values of $\Delta \varphi$. 

In order to see why it works, one must keep in mind that turbulent energy accumulates around regions where $k_\parallel \approx 0$. With the conventional spectral representation this occurs where $n + m/q(r) \approx 0$. One often refers to this occurrence as the "resonance condition". With the representation given by (\ref{k-par-sin}) the condition is
\begin{equation}
\label{k-par-sin-res-cond}
\sin [(n+m/q(r)) \Delta \varphi] \approx 0
\end{equation}
which is satisfied by all combinations $(n+m/q(r)) \Delta \varphi = k \pi$ with $k$ integer, up to some maximum value. It is easy to see that when $\Delta \theta << \Delta \varphi$, the maximum allowed value of $m$, $\pi/\Delta \theta$, is sufficiently large that the resonance condition occurs for several values of $k$.  
As a consequence, for a given domain in $(n,m)$ space, several bands satisfying (\ref{k-par-sin-res-cond}) are populated. The number of bands increases with the ratio $\Delta \varphi/\Delta \theta$, while the total number of sites in $(n,m)$ space that satisfy (\ref{k-par-sin-res-cond}) stays roughly constant.

\begin{figure}
\centering
\includegraphics[width=8cm]{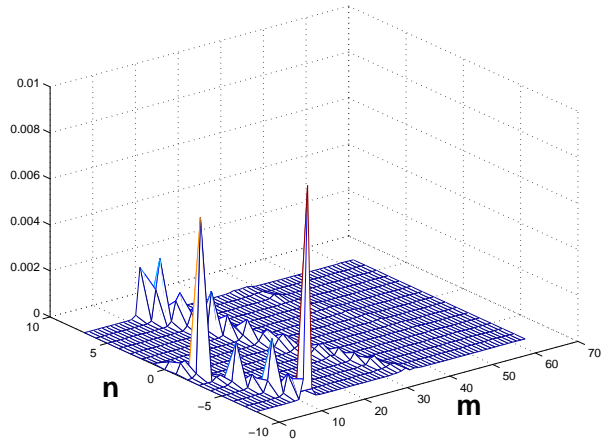}
\caption{Temperature fluctuations spectrum in field-aligned coordinates} 
\label{fig-spectrum-red}
\end{figure}

The situation is illustrated in Fig.~\ref{fig-spectrum-red}, which shows the spectrum of the same field as in Fig.~\ref{fig-spectrum}, after a transformation
to real space, reduction of the toroidal resolution by a factor 8, and transformed back to mode number space.
This result can be seen as the effect of aliasing, which is beneficial in this instance, since aliased modes occupy formerly empty space. In this case one can say that aliasing acts as an efficient mean of data compression.    

When solving a plasma micro-turbulence model, the condition of small parallel wave-number is produced dynamically. If the initial conditions are such that the modal composition satisfies the appropriate resonance condition, this property is preserved by the subsequent evolution. In particular, initial conditions with a spectrum like the one shown in Fig.~\ref{fig-spectrum-red} will continue to have a spectrum of this type if the parallel derivative operator is of the form given in (\ref{k-par-sin}).

It has then been a simple matter to convert a code like ETAI3D to field aligned coordinates by replacing all the occurrences of the combination of $n+m/q(r)$ by expression (\ref{k-par-sin}). An important point to note is that the spectral nature of the code allows one to keep the Landau closure on the parallel heat flux in the original form proportional to $|\gradpar|$ as given in Eq.~\ref{T}. It is so also for the modified version with a coarser toroidal resolution, since $|\gradpar|$ is replaced by the absolute value of (\ref{k-par-sin}). In this respect the new code is unique, since parallel derivatives in field-aligned codes are usually treated with finite differences in real space. This makes it difficult to deal with nonlocal dissipation operators proportional to $|\gradpar|$, which are then replaced by higher derivatives. It is important to keep in mind this peculiarity when examining the results of the tests of the new version of the code, presented in the following section.

\section{Tests \label{tests}}

Tests of the new version of ETAI3D were carried out at values of $\rho_{\displaystyle *}$ of $1/50$ and $1/100$ in cylindrical mode (with the curvature operator $\omega_d$ set to zero) and in toroidal mode, with various initial conditions. The tests consist in comparing the results by varying the number of toroidal points $N_\varphi$. All the tests were carried out with a simulation domain comprised between $r=0.5$ and $r=1$, and with a safety factor profile $q = 2r$.

\begin{figure}[h]
\centering
\includegraphics[width=8cm]{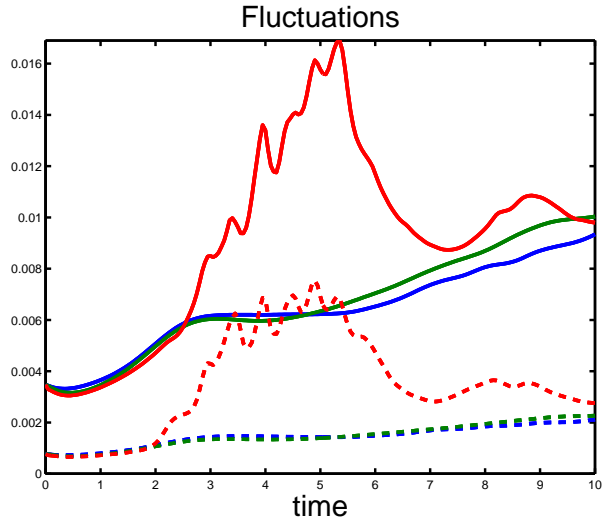}
\caption{Comparison of the turbulent fluctuation amplitude of potential (solid) and temperature (dash), obtained at $N_\varphi=128$ (blue lines), $N_\varphi=32$ (green lines) and $N_\varphi=16$ (red lines)}
\label{fig-test-cyl-50}
\end{figure}

In general, similar results were obtained for all values of $N_\varphi$ as low as $32$. On observes that the behaviour at the lower value of $N_\varphi = 16$ can be very different, at least transiently. The situation is illustrated in Fig.~\ref{fig-test-cyl-50}. The plots show the amplitude of turbulent fluctuations of a cylindrical case with $\rhost=1/50$, obtained at $N_\theta = 128$ (fixed) and with $N_\varphi$ of $128$, $32$, and $16$. Initial conditions are those of an unstable temperature profile with the addition of small fluctuations that satisfy the small parallel gradient condition. The system initially evolves linearly, with the fluctuations growing exponentially, and eventually saturates in a turbulent state.  

The blue lines represent simulations at $N_\varphi=128$. When compared with the former version of the code that does not employ field-aligned coordinates, the time traces are virtually undistinguishable (these are not shown here). 
This is taken as the reference case. As the number of toroidal points is reduced to $N_\varphi=32$ (green lines) one can note some difference in the nonlinear phase. This can be attributed to the reduced accuracy inherent in using a larger $\Delta \varphi$ to compute the parallel derivative.
Upon a further reduction to $N_\varphi=16$ (red lines), one observes that the traces strongly overshoot the reference case when entering the nonlinear phase, although eventually one observes signs of recovery.    

\begin{figure}
\centering
\includegraphics[width=8cm]{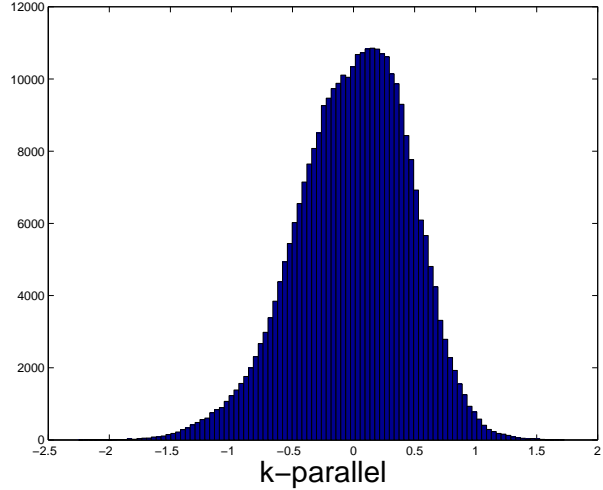}
\caption{Histogram of $k_\parallel$ for a $N_\varphi=32$ case}
\label{fig-k-par-dist-32}
\end{figure}

\begin{figure}
\centering
\includegraphics[width=8cm]{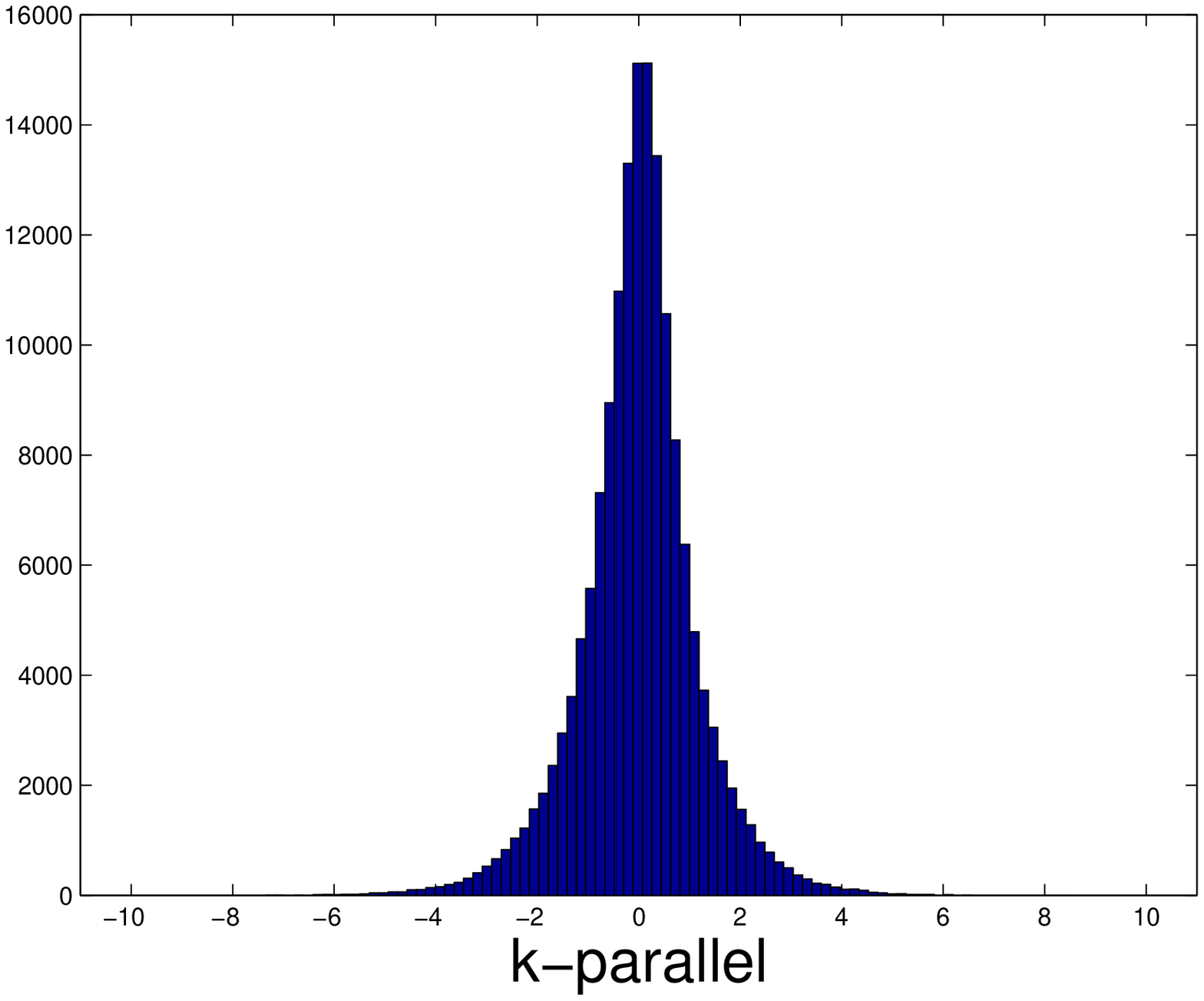}
\caption{Histogram of $k_\parallel$ for a $N_\varphi=16$ case }
\label{fig-k-par-dist-16}
\end{figure}

\begin{figure}
\centering
\includegraphics[width=8cm]{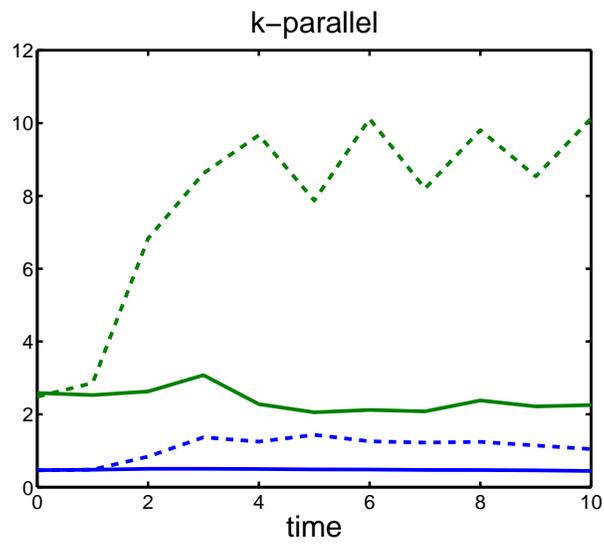}
\caption{Maximum values (dashed lines) and r.m.s. values (solid lines) of $k_\parallel$ as a function of time, for the two cases at $N_\varphi = 16$ (green upper lines) and $N_\varphi = 32$ (blue lower lines).}
\label{fig-k-par-vs-time}
\end{figure}

In order to understand why the code breaks down, consider that the following criterion must be satisfied at every point in the simulation domain
\begin{equation}
\label{k-par-criterion}
k_\parallel(r,\theta,\varphi) \Delta \varphi < 1 
\end{equation}
where $k_\parallel(r,\theta,\varphi)$ is a local measure of the parallel gradient. The following definition is employed for any function $F(r,\theta,\varphi)$
\begin{equation}
\label{k-par-def}
k_\parallel(r,\theta,\varphi) = \nabla_{\parallel} F / ||F||
\end{equation}
where $||F||$ is the mean squared norm at radial location $r$, 
$$
||F|| \equiv 
\left [ \frac{1}{(2 \pi)^2} \int d \theta d \varphi F^2 \right ]^{1/2}.
$$
Fig.~\ref{fig-k-par-dist-32} and Fig.~\ref{fig-k-par-dist-16} show the distribution function of $k_\parallel$ of the $N_\varphi=32$ and $N_\varphi=16$ cases respectively, at the end of the simulation. One notices that the distribution of the $N_\varphi=16$ case is broader, with "wings" at fairly high $k_\parallel$. Fig.~\ref{fig-k-par-vs-time} compares the maximum and the r.m.s. value of $k_\parallel$ in the two cases. It is apparent that criterion (\ref{k-par-criterion}) is satisfied in the $N_\varphi=32$ case, since the highest value of $k_\parallel$ is about $3$ so that the maximum value of expression (\ref{k-par-criterion}) achieved at any point in the simulation domain is about $0.5$. It is also clear that when $N_\varphi$ is reduced to $16$, the criterion is violated. When this happens, the discretised system is not able to adequately represent the continuum system. 

Several tests show that the linear phase is the easiest to simulate (the one that tolerates smaller $N_\varphi$), while the most critical one is the transient phase after the linear one when the system adjusts to the saturated turbulence. 
In this phase, locally higher-than-expected values of $k_\parallel$ can occur, although the typical value stays close to the inverse of the characteristic system length.

\begin{figure}
\centering
  \includegraphics[width=8cm]{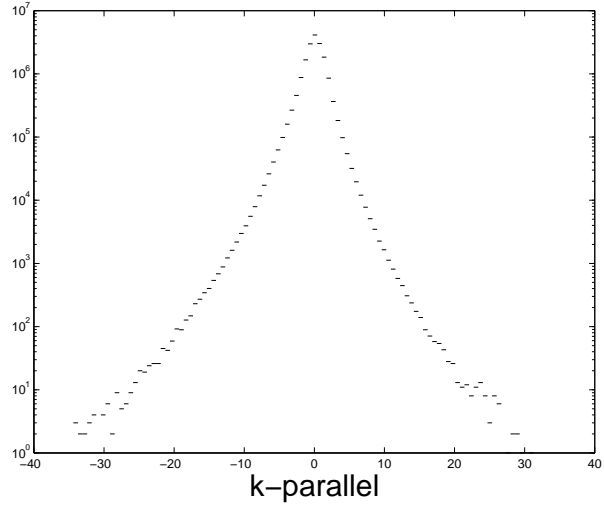}
\caption{Histogram of $k_\parallel$ (vorticity) for a Cyclone case}
\label{fig-k-par-Cyclone}
\end{figure}

It must be understood that the generation of locally strong parallel gradients depends on the model, not on the method. Indeed, analysis of older simulations carried out with the conventional uniform discretisation in the angles, also show this behaviour. An example drawn from data from the European code benchmarking~\cite{EUbench} on Cyclone test cases~\cite{Cyclone} is given in Fig.~\ref{fig-k-par-Cyclone}. In this instance, values of $k_\parallel$ as high as $40$ (although with a minuscule probability) can occur. This would require a minimum of $N_\varphi \approx 240$ to stay on the safe side. 

For any given plasma turbulence model it is obviously important to assess how the maximum parallel gradient, that is generated dynamically, depends on the parameters. The interest of the field-alignment technique would be somewhat reduced if one were to discover, for example, that, for a given class of models, the maximum parallel gradient increases when the scale separation parameter $\rho_{\displaystyle *}$ decreases. One must also stress that this behaviour may be model dependent. In particular, in the case of fluid models, it may depend on the parallel closure. A Landau closure as the one employed here is somewhat weaker that a diffusive-type closure and it could allow a more frequent development of fine structures in the parallel direction. It would also be extremely interesting to study the behaviour of a kinetic model in this respect. 

\section{Conclusions \label{conclusions}}

The main results of this work are here summarised. 
Field aligned techniques are necessary to optimise plasma turbulence codes. This work has shown that there are only two practical ways to achieve this. In the one employed by most codes, the position along the field lines is labelled by the poloidal angle like in the ballooning approximation of linear theory. In the one pursued here, points along field lines are labelled by the toroidal angle. 

This results in a more natural discretisation of the fields that has several advantages. In particular, it can deal with X-point geometries and both closed and open field lines.
Furthermore, as shown in the Appendix, the description in the poloidal plane, where perpendicular derivatives are computed in most situations, need not employ magnetic coordinates. Information on field lines is needed only to compute parallel derivatives. This opens the way to generic and efficient coding for global tokamak simulations. 

It has also been shown how conventional codes that employ a spectral, uniform, representation on a magnetic surface can be converted to the new coordinate system with minimal programming. This has been done for the ETAI3D code, which now exists in two versions
that solve exactly the same model equations without further approximations.

The comparison between the two versions has uncovered an unexpected property of the plasma turbulence model used for testing purposes. It has indeed been shown that parallel gradients can be intermittently higher than expected in the nonlinear phase. 
Parallel gradients are limited by the parallel heat flux, which, in this model, takes the form of a Landau closure. Landau closures are often used by gyro-fluid models. Then the key question, which cannot be answered here, is whether intermittently high parallel gradients are a real property of the weakly collisional plasma turbulence or an indication that the closure on the parallel heat flux is not adequate enough. Obviously, the whole construct of field-alignment rely on uniformly small parallel gradients for an efficient implementation.
 
\section*{Acknowledgments}

This work was partly motivated and has greatly benefitted from stimulating discussions with B. Scott, mostly during working sessions of the EFDA Integrated Tokamak Modelling Task Force. The author would also like to thank D. Escande, G. Falchetto, X. Garbet, G. Hammett and F. Schwander for useful comments and suggestions.

This work, supported by the European Communities under the contract of Association between EURATOM and CEA, was carried out within the framework of the European Fusion Development Agreement. The views and opinions expressed herein do not necessarily reflect those of the European Commission.
%

\appendix
\section{Appendix}

This appendix sketches how the approach to field-aligned coordinates described in this work can be extended to avoid the use of magnetic surface coordinates to discretise the fields. The fields can then be discretised on a given grid related to the laboratory (or machine) reference frame. For a tokamak, these are the usual $(R,\Phi,Z)$ cylindrical coordinates such that $Z$ is the direction of the torus axis, $R$ the distance from the axis and $\Phi$ the toroidal angle.
The method is built on two concepts developed here: that micro-turbulent fields need to be known with high resolution only in two directions, and that knowledge of the magnetic field structure is needed only to compute parallel derivatives, which can be obtained by finite differences such that values at end points are obtained by interpolation.

In the following, one considers a simple static, low-$\beta$, cylindrical equilibrium, such that the suitably normalised magnetic field is given by
\BE
\vB = \vb(\vx) + \vuz
\label{mag-field}
\EE
where one employs a Cartesian reference system $(x,y,z)$ such that $\vuz$ is the direction of the magnetic axis, the main magnetic field along $z$ is constant and normalised to unity, and $\vb(\vx)$ is the poloidal magnetic field in the poloidal plane $(x,y)$. The vector $\vx$ indicates the position in this plane. The poloidal field can be written in terms of a flux function $\psi(\vx)$ such that
\BE
\vb = \vnabla \times (\psi \vuz)   
\label{pol-field}
\EE      
Magnetic surfaces can be labelled by the value of $\psi$. Both closed and open field lines can be treated.
The parallel derivative operator is given by
\BE
\gradpar = \vb \cdot \vnabla + \partial / \partial z
\label{gradpar-gen}
\EE
One has to look for a change of coordinates from the original $(x,y,z)$ to a new set ($\xi^{\alpha},s$) such that $s$ can be treated as a slowly-varying coordinate and only the two $\xi^{\alpha}$'s ($\alpha = 1,2$) carry the information on the small scales. Taking advantage from what was learnt in Sec.~\ref{new-approach}, one considers a transformation of the form:
\BEA
&& \xi^\alpha = V^\alpha(\vx) + C^\alpha (\vx) (z - z_k) \label{gen-transf1}
\\
&& s = z - z_k \label{gen-transf2}
\EEA
where $k$ labels a given sector in the $z$ direction and $V^\alpha(\vx)$ and $C^\alpha (\vx)$ are yet unknown functions.
In terms of the new variables the parallel derivative is given by
\BE
\label{par-der-gen-transf}
\gradpar = b^\alpha \frac{ \partial V^\beta}{\partial x^\alpha} \frac{\partial}{\partial \xi^\beta}
+ 
(z-z_k) b^\alpha \frac{ \partial C^\beta}{\partial x^\alpha} \frac{\partial}{\partial \xi^\beta}
+ 
C^\beta \frac{\partial}{\partial \xi^\beta}
+
\frac{\partial}{\partial s}
\EE
In order to eliminate the fast-varying derivatives one has to satisfy the conditions:
\BEA
&& C^\alpha = - b^\beta \frac{ \partial V^\alpha}{\partial x^\beta} \label{cond1}
\\
&& b^\alpha \frac{ \partial C^\beta}{\partial x^\alpha} = 0
\label{cond2}
\EEA

In the following, the Poisson bracket notation is used such that the operation
$b^\alpha \frac{ \partial A} {\partial x^\alpha} \equiv [\psi,A]$ for any function $A$. Eqs.~(\ref{cond1}-\ref{cond2}) can then be written as
\BE
\label{cond-V}
[\psi,[\psi,V^\alpha]] = 0
\EE
Consider a function $\chi(x,y,)$ such that $[\psi,\chi]=1$, whose solution can be found with the method of characteristics. This function identifies the position on $\psi={\rm const}$ surfaces and plays the role of a poloidal angle.
Then a general solution of Eq.~\ref{cond-V} is
\BE
\label{gen-sol-V}
V^\alpha = f^\alpha(\psi) + g^\alpha(\psi) \chi(x,y),
\EE 
where $f^\alpha$ and $g^\alpha$ are arbitrary functions.
Thus, suitable solutions to Eqs.~(\ref{cond1}-\ref{cond2}) are found, such that the parallel gradient reduces to 
$$
\gradpar = \frac{\partial} {\partial s} ,
$$
computed at fixed $\xi^\alpha$.

In an actual code implementation based on finite differences, one defines any field at nodes in the Cartesian $(x,y,z)$ grid. Derivatives in the poloidal plane are computed in this reference frame by holding $z$ constant. 

In order to compute the parallel derivative by finite differences, one has to use function values at points $(\vx + \Delta \vx,z_k + \Delta z)$ corresponding to a given increment $\Delta s$ along s. This means finding end points along field lines for a given displacement $\Delta z$ along $z$. From Eqs.~(\ref{gen-transf1}-\ref{gen-transf2}) one finds the {\sl finite difference} equations for the unknown increments $\Delta \vx$
\BE
\label{FD-equation}
[f^\alpha(\psi) + g^\alpha(\psi) \chi(\vx)]_{\vx + \Delta \vx} -
[g^\alpha(\psi)]_{\vx} \Delta z = 
[f^\alpha(\psi) + g^\alpha(\psi) \chi(\vx)]_{\vx}
\EE
Solutions to these equations exist such as $\psi(\vx + \Delta \vx) =\psi(\vx)$ and $\chi(\vx + \Delta \vx) =\chi(\vx) + \Delta z$. 
Thus end points for FD computations are obtained by moving along field lines for a given increment along $z$.

Finally, in the special case of circular flux surfaces, such that $\psi = \psi(x^2+y^2)$, one finds that $\chi$ is proportional to the usual poloidal angle, via the safety factor. Then the scheme sketched here reduces to the one described in Sec.~\ref{new-approach}.

\end{document}